# DEVELOPMENT OF REBUNCHING CAVITIES AT IAP*

C. P. Welsch, K.-U. Kühnel, A. Schempp
Institut für Angewandte Physik
*Johann Wolfgang Goethe Universität, Frankfurt am Main, Germany*

*Abstract*

A focus of work at IAP has been the development and optimization of spiral loaded cavities since the 1970s [1]. These cavities feature a high efficiency, a compact design and a big variety of possible fields of application.
They find use both as bunchers and post accelerators to vary the final energy of the beam. In comparison to other available designs, the advantage of these structures lies in their small size. Furthermore they can easily be tuned to the required resonance frequency by varying the length of the spiral. Due to the small size of the cavities the required budget can also be kept low.
Here, two slightly different types of spiral loaded cavities, which were built for the REX-ISOLDE project at CERN and the intensity upgrade program at GSI are being discussed.

## 1 COMMON FEATURES

### 1.1 General Remarks

The variation of the final energy of accelerators cannot be optained with long linear accelerators like alvarez, wideroe or IH structures alone. Therefore, short additional resonators are very often used. They allow changing the energy of the particles as well as the velocity distribution of the beam.

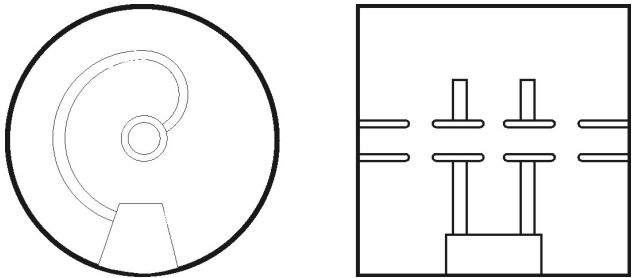

Figure 1: Schematic Drawing of spiral loaded cavity

In a cylindrical external tank one or more spiral arms are placed. At their open end each spiral carries a drift tube running along the symmetry axis of the cavity and thus forming several accelerating gaps.

Different designs of the spiral arms are possible, depending on the required resonance frequency. With Splitring structures resonance frequencies of up to 300 MHz can be achieved, whereas archimedic type spirals find their application at resonance frequencies as low as 27 MHz [2].

### 1.2 Characteristic Parameters

The transittime factor $T$ indicates the flexibility of an accelerator as the quotient of the effectively experienced voltage and the total gap voltage. The $R_p$-value is defined as the effective cavity voltage squared over the power $P$ put into the cavity.

$$T = \frac{U_{eff,Gap}}{U_{tot}} \qquad R_p = \frac{U^2_{eff,Cav.}}{P}$$

With $n$ as the number of gaps, the $R_p$-value is proportional to $n^2$ in case of the single-spiral structure and proportional to $n$ for a splitring structure.
The transittime factor is the product of a velocity dependent and a geometry based term [1]

$$T_G = \frac{\beta \lambda}{\pi(d + 0.85 \cdot r)} \cdot sin\left(\frac{\pi(0.85 \cdot r)}{\beta \lambda}\right) \cdot \frac{1}{I_0\left(\frac{2\pi a}{\beta \lambda}\right)}$$

$$T_V = \frac{2 \cdot \beta}{n \cdot \pi \cdot (\beta - \beta_0)} \cdot sin\left(\frac{n \cdot \pi \cdot (\beta - \beta_0)}{2 \cdot \beta}\right)$$

where  $d$ = width of acceleration gap
  $r$ = radius of curvature of drifttubes ($r_{out}$-$r_{in}$)
  $a$ = radius of aperture
  $\beta_0$ = $v_0/c$;  $\beta = v/c$
  $v_0/v$ = reference / real velocity of particles

Fig. 2 shows a plot of the velocity dependent term in case of the 36 MHz 11.4 MeV/u rebuncher built for GSI.

The wide spectrum of possible applications is clearly shown in this plot. While many-gap structures are limited to a narrow energy interval, a 2- or 3-gap accelerator can handle even particles with a large deviation from the design energy. This flexibility is characteristic for all of the here presented structures.

* Work supported by the BMBF

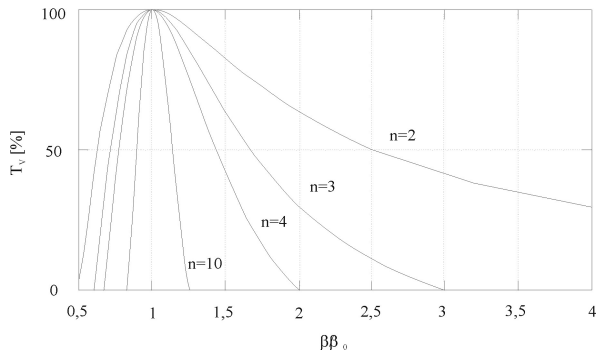

Figure 2: velocity depend term of the transittime factor

The resonance frequency of the structures can also be determined from the geometrical dimensions alone [2,3]. It is possible to describe a spiral loaded cavity as a transission line loaded with capacitors or as a set of capacitors and inductors. Their values can be calculated analytically with a high accuracy. Fig. 3 shows the calculated capacity $C$ at each $mm$ of the 36 MHz 11.4 MeV/u spiral in the transfer channel at GSI.

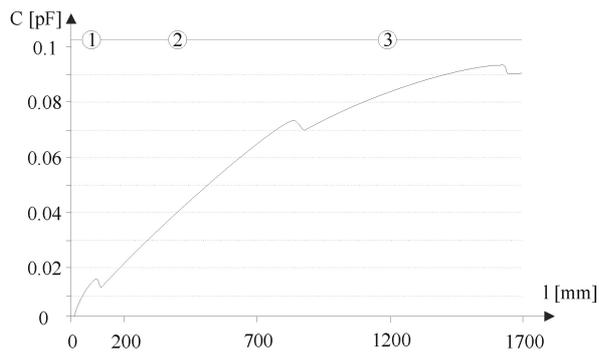

Figure 3: capacity $C$ between spiral arm and tank

In addition to these analytical calculations, all of the above parameters, the quality factor $Q$ and the shunt impedance $\eta$ are determined with the MAFIA program. Using the obtained field distributions, optimisation of the structures is done.

## 2 THE INTENSITY UPGRADE PROGRAM AT GSI

It is the goal of the intensity upgrade program at GSI to fill the SIS up to its space charge limit. This requires a beam current of up to 15 emA of $^{238}U^{4+}$ ions. Therefore the old Wideroe structure of the UNILAC has been replaced by an RFQ and two IH-structures, especially designed to accelerate high intensity, low charge state beams.
Special attention had to be paid in the stripping section at 1.4MeV/u, where high space charge effects are present and focussing elements in both transverse and longitudinal direction are necessary.

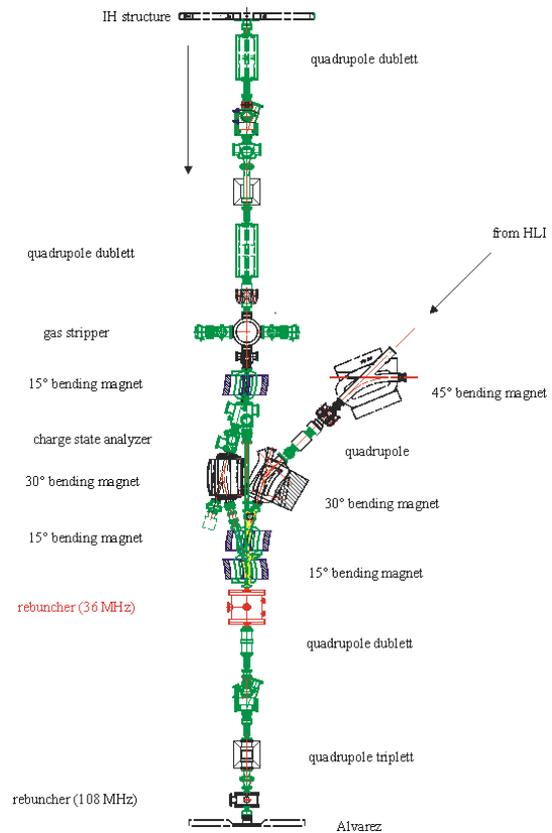

Figure 4: Stripper section of the UNILAC

Even though an 108 MHz Rebuncher was already present in the section in front of the Alvarez-linac, a stronger rebuncher was necessary to counteract the longitudinal defocusing effects. The resonance frequency was chosen to be 36 MHz, like the frequency in the new HSI, providing a larger region of linear forces.

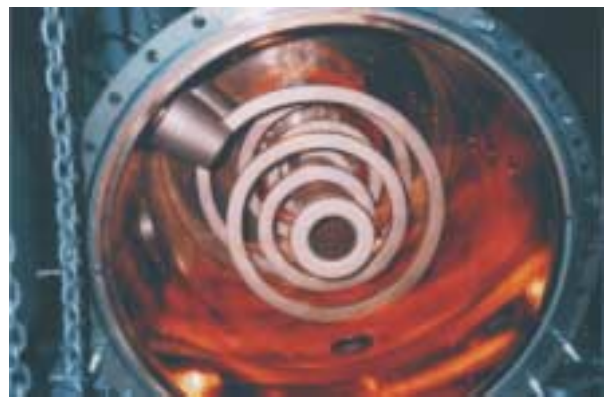

Figure 5: View inside 36 MHz 1.4 MeV/u rebuncher

This low frequency and the limited space available required a very compact structure. A two-spiral-cavity with constant pitch ("archimedic type") was chosen, as shown in Fig. 5. Mechanical stability was improved by adding two spiral tubings axially aside each spiral

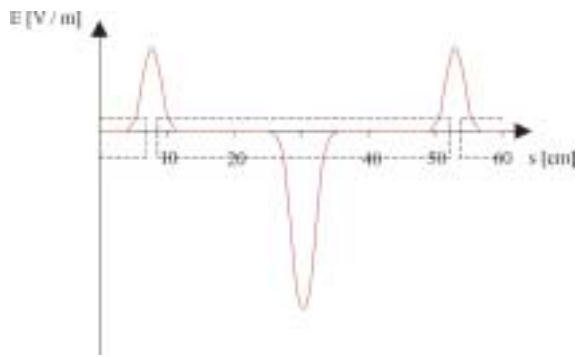

Figure 6: distribution of electric field along axis

A number of simulations was done to optimize the particle motion through the buncher [3,4]; always taking into account the various kinds of different ions that are going to be used at GSI.

Fig. 6 shows the distribution of the electric field along the beam axis. Since the gap in the middle of the structure is as wide as the outer ones, its field shows twice the amplitude. The rebuncher resonators are now integrated into the beamline in routine operation.

## 3 REBUNCHER FOR REX-ISOLDE

REX-ISOLDE (Radioactive Beam Experiment at ISOLDE) is a new project at the online mass separator ISOLDE / CERN [5].

The structure of exotic and neutron rich nuclei, especially Na, Mg, K and Ca, will be studied. The nuclei have to reach energies in the order of the coulomb barrier.

By measuring the arising γ-rays both, static and dynamic properties of these exotic nuclei can be examined.

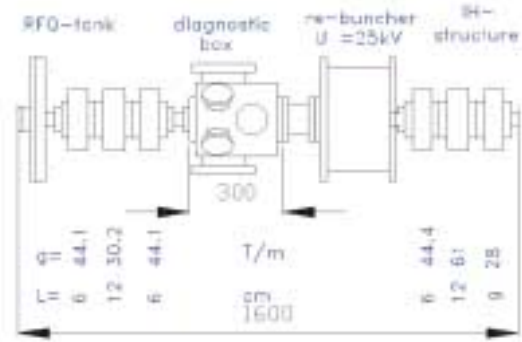

Figure 7: Part of the REX-ISOLDE beamline

In order to accelerate the radioactive beam to the necessary energy, a combination of an ion source, penning trap, charge state breeder, an RFQ, an IH-structure and a 7-gap resonator is used. The longitudinal match is achieved by inserting a rebuncher between RFQ and IH structure. A short high energy 7-gap spiral cavity allows an energy variation between 0.8 MeV/u and 2.2 MeV/u.

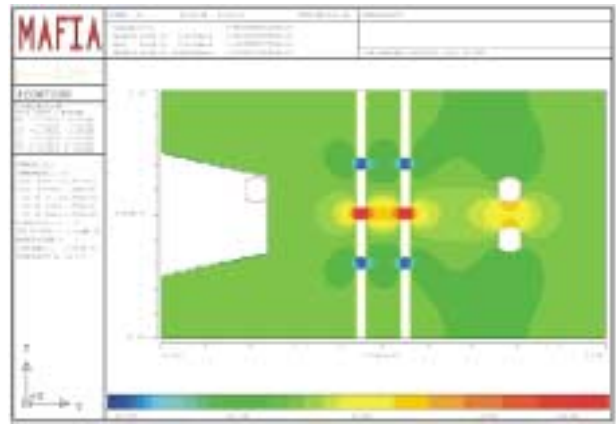

Figure 8: calculated field distribution

For the rebunching cavity the shunt impedance has been optimized at a high separation of the 0- and π-mode. The calculated temperature distribution indicated that additional cooling is necessary. An effective power of 2 kW is put into the structure, which was tested successfully in Munich and is now installed at REX ISOLDE.

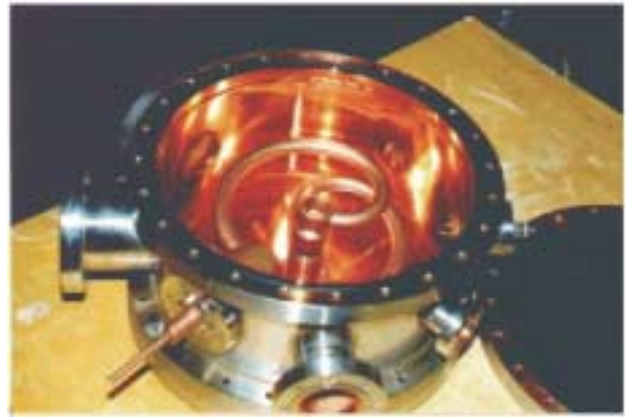

Figure 9: View into open splitring cavity

## REFERENCES

[1] A. Schempp et al, NIM 135, 409 (1976)
[2] J. Häuser, "Eigenschaften von Spiralresonatoren", thesis, University of Frankfurt, 1989
[3] C. Welsch "Aufbau und Untersuchung von Spiralresonatoren für den Hochstrominjektor der GSI", Diplomarbeit, University of Frankfurt, 1999, http://www.carsten-welsch.net
[4] W. Barth et al, "High Current Beam Dynamics for the Upgraded UNILAC", PAC,a Vancouver, 1997
[5] K.U. Kühnel „Aufbau eines Splitringresonators als Rebuncher für das REX-ISOLDE Projekt", Diplomarbeit, University of Frankfurt, 1999